\documentclass[runningheads]{llncs}
\usepackage[T1]{fontenc}
\usepackage{graphicx}
\usepackage{multirow}
\usepackage{tablefootnote}
\usepackage{amsmath}
\usepackage{caption} 
\begin{document}
\title{Evaluating Membership Inference Attacks in heterogeneous-data setups}
\titlerunning{Evaluating MIA in heterogeneous-data setups}
%
\author{Bram van Dartel\inst{1} \and
    Marc Damie\inst{1,2} \and
    Florian Hahn\inst{1}
}
\authorrunning{van Dartel et al.}
%
\institute{University of Twente, The Netherlands \and
    Inria, France}
\maketitle              
\begin{abstract}
    Among all privacy attacks against Machine Learning (ML), membership inference attacks (MIA) attracted the most attention.
    In these attacks, the attacker is given an ML model and a data point, and they must infer whether the data point was used for training.
    The attacker also has an auxiliary dataset to tune their inference algorithm.

    Attack papers commonly simulate setups in which the attacker's and the target's datasets are sampled from the same distribution.
    This setting is convenient to perform experiments, but it rarely holds in practice.
    ML literature commonly starts with similar simplifying assumptions (i.e., ``i.i.d.'' datasets), and later generalizes the results to support heterogeneous data distributions.
    Similarly, our work makes a first step in the generalization of the MIA evaluation to heterogeneous data.

    First, we design a metric to measure the heterogeneity between any pair of tabular data distributions.
    This metric provides a continuous scale to analyze the phenomenon.
    Second, we compare two methods to simulate a data heterogeneity between the target and the attacker.
    These setups provide opposite performances: 90\% attack accuracy vs. 50\% (i.e., random guessing).
    Our results show that the MIA accuracy depends on the experimental setup; and even if research on MIA considers heterogeneous data setups, we have no standardized baseline of how to simulate it.
    The lack of such a baseline for MIA experiments poses a significant challenge to risk assessments in real-world machine learning scenarios.
\end{abstract}
\section{Introduction}
Machine Learning (ML) has become an essential tool to process large amount of data.
Many applications involving personal data (e.g., in healthcare) have integrated ML-based solutions; raising concerns related to data privacy.
In particular, a lot of attention was drawn at the leakage of private information by ML models trained on personal data.

To study this privacy leakage, many papers \cite{du_sok_2024,fredrikson_model_2015,nasr_comprehensive_2019,shokri_membership_2017,zhao_loki_2024,zhu_deep_2019} proposed attacks extracting private information contained in ML models.
Among all attacks, Membership Inference Attacks (MIA) have a special role in this literature.

In these attacks, the attacker is given an ML model and a data point, and they must infer whether the point was in the training set.
To perform the attack, the attacker knows an auxiliary dataset commonly used to train a membership inference algorithm.
Several works \cite{chatzikokolakis_bayes_2023,liu_ml-doctor_2022} use MIA as the gold standard to measure the privacy of a model.
Chatzikokolakis et al. \cite{chatzikokolakis_bayes_2023} even built a novel privacy definition  based on membership inference.

The experimental setup used in attack papers makes some implicit assumptions.
In particular, they usually simulate attacks in which the attacker's auxiliary dataset and the target's dataset are sampled from the same distribution.
This implicit assumption is rarely discussed in existing papers, but has a major impact on the practicality of the attacks.
It is questionable, if this assumption does hold in real-world ML use cases \cite{humphries_investigating_2023}.
ML literature often uses the term of ``distribution shift'' \cite{chen_mandoline_2021} to refer to this divergence between two data distributions.

This phenomenon has a well-known symptom: heterogeneous datasets.
Data heterogeneity is a recurrent problem that multiple papers studied in privacy-preserving machine learning \cite{dennis_heterogeneity_2021,noble_differentially_2022}, especially in Federated Learning (FL).
Federated Learning is a popular ML paradigm \cite{kairouz_advances_2021} to train ML models on decentralized private data.
Recently, several privacy attacks \cite{du_sok_2024,nasr_comprehensive_2019,zhao_loki_2024} including MIA have been extended to this paradigm.
However, none of these works considered attack setups with data heterogeneity.
These works leave a key question open: what are the effects of a realistic data heterogeneity on MIA?

Humphries et al. \cite{humphries_investigating_2023} is the only work that studied MIA in heterogeneous-data setups.
In particular, they provide dedicated attack mitigations for this specific setup.
In a heterogeneous-data setup, they report attack accuracy up to 90\%, but our study \textbf{highlights some contradictory results}.
This contradiction comes from a different sampling method for the attacker's auxiliary dataset in the MIA simulations.

\paragraph{Our contributions}
\begin{enumerate}
    \item A metric to \textbf{estimate the heterogeneity} between tabular datasets.
    \item A new method to generate heterogeneous-data setups different from the method used by Humphries et al. \cite{humphries_investigating_2023}.
    \item A comparison of the two methods showing that they \textbf{lead to opposite results}: 90\% accuracy (for Humphries et al.'s setup) vs. 50\% (for the other).
\end{enumerate}

\paragraph{Focus}
Like related works \cite{humphries_investigating_2023,li_federated_2022}, we focus on classification datasets, because most attack papers \cite{fredrikson_model_2015,shokri_membership_2017,zhao_loki_2024,zhu_deep_2019} targeted classification models.
In particular, we use tabular datasets, which simplifies our fine-grained analysis of the impact of data heterogeneity.
We leave the extension to other types of datasets (e.g., images) for future work.

Classification datasets contain two components: the features and the label (also called the ``class'').
For example, the ``\texttt{Students}'' dataset \cite{cortez_using_2008} is a classic tabular dataset to analyze the student performance in secondary school.
In this dataset, the label is the student result (i.e., pass or fail), and the features include various information about the student (e.g., age, grades, etc.).
The notation $x$ usually refers to the feature vector and $y$ to the label.

\section{Data heterogeneity metric}

\subsection{Defining data heterogeneity}
Data heterogeneity is a complex concept with many concurrent definitions.
In the context of Federated Learning, Li et al. \cite{li_federated_2022} identified three types of data heterogeneity: quantity imbalance (i.e., a dataset is significantly larger than the other), label imbalance (i.e., one label being more represented in one dataset than in the other), and feature imbalance (i.e., the feature vectors are not sampled from the same data distribution in each dataset).

We argue that feature imbalance should be the standard focus for works on data heterogeneity in MIA.
On the one hand, quantity imbalance is not related to the data distribution.
On the other hand, label imbalance can relate to data distribution, but it ignores any heterogeneity that may occur in the features.

For example, we can have two cancer detection datasets: one from Asia and one from Europe.
If both dataset have the same label distribution (i.e., same proportion of cancer), it does not imply that both datasets have the same data distribution.
Most likely, some divergences should exist in the feature space because medical problems vary across populations.
This example highlights that \textbf{feature imbalance is a more reliable symptom of distribution shift}.

However, feature imbalance is harder to measure: there is no reference metric for it; contrary to label imbalance that have several known metrics \cite{gutierrez_fedartml_2024}.
Our first goal is then to provide a generic feature-imbalance metric for tabular data.

\subsection{Measuring the distribution shift}
Let $\mathcal{X}_\text{tgt}$ (resp. $\mathcal{X}_\text{atk}$) be the distribution from which the target's (resp. attacker's) dataset is sampled.
We want to measure the divergence between $\mathcal{X}_\text{tgt}$ and $\mathcal{X}_\text{atk}$.

Statistical distances are tools designed for this purpose: they measure the distance between two probability distributions.
The statistics literature presents many statistical distances \cite{deza_encyclopedia_2009,venturini_statistical_2015}, each with advantages and disadvantages.

Unfortunately, these metrics cannot be used naively on our distributions $\mathcal{X}_\text{tgt}$ and $\mathcal{X}_\text{atk}$ for two reasons.
First, classification datasets have by definition one feature distribution per class.
Otherwise, the classes would be impossible to distinguish.
Second, statistical distances have closed-form formula for reference distributions (e.g., Gaussian distributions), but their computational cost is exponential (with the number of dimensions) for generic distributions.

\begin{figure}
    \centering
    \includegraphics[width=0.5\linewidth]{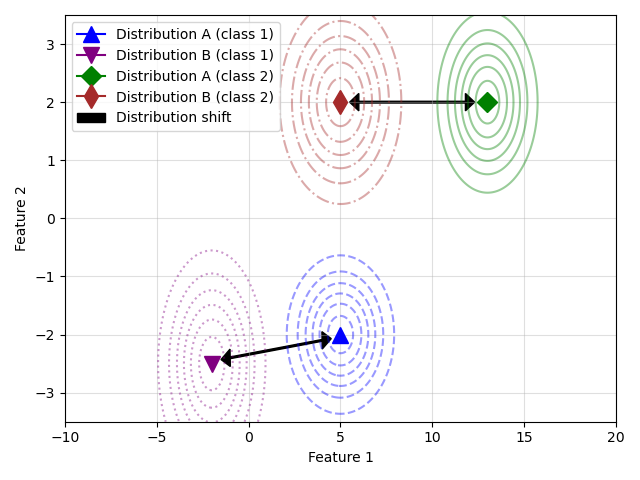}
    \caption{\emph{Simplistic} example of a ``per-class'' distribution shift for data distributions with two classes.}
    \label{fig:distribution-shift}
\end{figure}

We can extend easily statistical distances to take into account the classes.
Let us consider two classes $1$ and $2$, and the definition naturally generalizes to $K$ classes.
We have four distributions: $(\mathcal{X}^{(1)}_{\text{tgt}}, \mathcal{X}^{(2)}_{\text{tgt}})$ and $(\mathcal{X}^{(1)}_{\text{atk}}, \mathcal{X}^{(2)}_{\text{atk}})$.
Let $d(\mathcal{X}, \mathcal{X}')$ be any (standard) statistical distance between two probability distributions.
We can build a ``multi-class metric'': $d_\text{multi}(\mathcal{X}_{\text{tgt}}, \mathcal{X}_{\text{atk}}) = \frac{1}{2} (d(\mathcal{X}^{(1)}_{\text{tgt}}, \mathcal{X}^{(1)}_{\text{atk}}) + d(\mathcal{X}^{(2)}_{\text{tgt}}, \mathcal{X}^{(2)}_{\text{atk}}))$.
In other words, we compute the distance for each class and then average the results.
This multi-class distinction also allows ignoring any label imbalance in the data to focus solely on the feature imbalance.
Figure \ref{fig:distribution-shift} illustrates the ``per-class'' distribution shift captured by our metric.

To have a computationally efficient metric, we need an algorithm to ``transform'' any distribution into a multivariate Gaussian distribution, because statistical distances usually have closed-form formula for such distributions.
Multivariate Gaussians have two parameters: a mean vector and a covariance matrix.
For any distribution $\mathcal{X}_{*}$, we can define a Gaussian distribution $\widetilde{\mathcal{X}}_{*}$ such that $\text{Covariance}(\mathcal{X}_{*}) = \text{Covariance}(\widetilde{\mathcal{X}}_{*})$ and $\text{Mean}(\mathcal{X}_{*}) = \text{Mean}(\widetilde{\mathcal{X}}_{*})$.
To define $\widetilde{\mathcal{X}}_{*}$, we need to extract the covariance and mean information from $\mathcal{X}_{*}$.
This information can be estimated using the datasets given for a specific scenario under study.

This Gaussian distribution is a ``proxy'' in our metric: $d_\text{generic}(\mathcal{X}_{a}, \mathcal{X}_{b}) = d_\text{Gauss}(\widetilde{\mathcal{X}}_{a}, \widetilde{\mathcal{X}}_{b})$.
Instead of computing the distance between two distributions $\mathcal{X}_{a}$ and $\mathcal{X}_{b}$, we compute the distance between their proxy Gaussian distributions $\widetilde{\mathcal{X}}_{a}$ and $\widetilde{\mathcal{X}}_{b}$.
This metric $d_\text{generic}$ is not the real distance between $\mathcal{X}_{a}$ and $\mathcal{X}_{b}$, but it provides sensible approximation that can be computed efficiently.

Finally, we combine our multi-class transformation and the proxy distribution to build our data heterogeneity metric:
$$D(\mathcal{X}_{\text{tgt}}, \mathcal{X}_{\text{atk}}) = \frac{1}{2} (d(\widetilde{\mathcal{X}}^{(2)}_{\text{tgt}}, \widetilde{\mathcal{X}}^{(1)}_{\text{atk}}) + d(\widetilde{\mathcal{X}}^{(2)}_{\text{tgt}}, \widetilde{\mathcal{X}}^{(2)}_{\text{atk}}))$$

As base statistical distance $d$, we use the ``2-Wasserstein'' distance; a popular metric with a closed-formula for Gaussian distributions \cite{delon_gromovwasserstein_2022}.
This metric has value in $[0, \infty)$: the higher the value is, the more heterogeneous the distributions are.
Our approach works with any statistical distance having a closed-form formula for Gaussian distributions.

\paragraph{Beyond tabular data}
Our metric can be computed on any datasets, even images or time series.
However, this metric seems more adapted to tabular datasets.
Indeed, image (or time series) datasets have special structures.
For example, an image has pixels as features, and consecutive pixels are usually strongly correlated.
These datasets would deserve a dedicated metric integrating their specific properties.
Thus, we present our metric on tabular datasets, and recommend further research to extend it to other data formats.

\section{MIA in heterogeneous-data setups}
\subsection{Experimental setup}
As we motivate our work based on heterogeneous-data setups, we study the MIA described by Nasr et al. \cite{nasr_comprehensive_2019} because it focuses on Federated Learning setup.
We implemented their MIA using the Flower framework \cite{beutel_flower_2022}.
Our source code is available here: \url{https://anonymous.4open.science/r/MIA-IFL-096F/}

This attack uses a classic approach in MIA: the attacker trains a ``shadow'' model (using an auxiliary dataset) able to distinguish whether a point $(x,y)$ was used to train a model $\theta$.

Like \cite{humphries_investigating_2023}, we use two real-world datasets: \texttt{Students} and \texttt{Heart}.
The \texttt{Students} dataset \cite{cortez_using_2008} consists of around 650 student achievements in secondary education of two distinct Portuguese schools.
Like \cite{humphries_investigating_2023}, we exclude the intermediate grades because of their high correlation with the final grade.
The ML task is to predict whether a student passed or failed the course based on these features.
The \texttt{Heart} dataset  \cite{detrano_international_1989} consists of 900 data samples from four hospitals across the world: Long Beach (VA), Switzerland (CH), Cleveland (CL), and Hungary (HU).
The ML task is to predict the presence of a heart disease.

These two datasets are valuable for ML research because they provide a ``natural'' data heterogeneity (as illustrated in Table \ref{tab:results-our-splitting}).
For example, the ``\texttt{Heart}'' dataset includes data from four hospitals; each with a distinct distribution.
Our experiments rely on this natural heterogeneity to study MIAs under \emph{realistic} data heterogeneity.

\subsection{Results}
\paragraph{Dataset splitting}
To simulate an MIA, we must split our dataset into three disjoint subsets: the attacker's dataset, the target's (or training) dataset, and the ``non-members''.
The attacker uses their dataset to train their shadow model.
The target uses their dataset to train the \emph{target model}.
Finally, we build a ``challenge'' dataset with 50\% of training data (i.e., the members), and 50\% of ``non-members.''
The attack accuracy is computed based on the challenge dataset.

In classic MIA works (e.g.,\cite{nasr_comprehensive_2019}), the dataset is split uniformly at random.
This provides no data heterogeneity because all subsets would have the same distribution.
To simulate a heterogeneous-data setup, we rely on the ``natural splitting'' existing in the \texttt{Students} and \texttt{Heart} datasets.
For example, we provide the data from hospital VA to the target, and the data from hospital CH to the attacker.
We also keep a small subset of hospital VA to build our non-member dataset.
We can perform a similar ``natural splitting'' on the \texttt{Students} dataset.

\paragraph{Heterogeneous vs. uniform splitting}
Table \ref{tab:results-our-splitting} compares the uniform splitting (i.e., no heterogeneity) to the ``natural'' splitting on the \texttt{Students} and \texttt{Heart} datasets.
First, using our heterogeneity metric, we observe that the natural splitting induces a much higher data heterogeneity; e.g., $10^{4}$ vs. $10^{2}$ on \texttt{Students}.

The difference of heterogeneity produced by the natural and the uniform splittings shows that \texttt{Students} (same for \texttt{Heart}) is composed of several heterogeneous distributions.
If all the data from \texttt{Students} was drawn from the same distribution, the natural splitting and the uniform splitting would induce the same data heterogeneity.

Note that our metric is not equal to zero on the uniformly split dataset, while it generates homogeneous distributions.
Our metric relies on the \emph{estimation} of the distribution covariance and mean.
This statistical estimation induces noise making the metric not null, even for homogeneous data distributions.
However, we expect the metric to converge towards 0 when the dataset sizes increase (because the estimation noise would decrease).

\begin{table}[t]
    \centering
    \begin{tabular}{|c|c|c|c|}
        \hline
        Dataset                            & Dataset splitting & Heterogeneity       & Average Accuracy \\
        \hline
        \hline
        \multirow{2}{*}{\texttt{Students}} & Natural           & $2.19\times10^4$    & 50.30            \\
                                           & Uniform           & $1.97\times10^2$    & 57.11            \\
        \hline
        \multirow{2}{*}{\texttt{Heart}}    & Natural           & $1.84\times10^{41}$ & 47.75            \\
                                           & Uniform           & $1.14\times10^{10}$ & 51.56            \\\hline 
    \end{tabular}
    \caption{Data heterogeneity (between the attacker's and target's datasets) and MIA accuracy using two splitting methods.}
    \label{tab:results-our-splitting}
\end{table}

As the challenge dataset is balanced (50\% of members/non-members), the random guess has 50\% accuracy.
On the one hand, we observe that the uniform split (i.e., no heterogeneity) provides a slightly higher accuracy than the natural split on \texttt{Students}: 57\% vs. 50\% accuracy.
On the other hand, both splitting methods provide an accuracy close to 50\% on \texttt{Heart} (i.e., inefficient MIA).
Overall, all these results highlights low attack accuracy.

Based on the existing results of Humphries et al. \cite{humphries_investigating_2023}, this low accuracy is surprising.
In a heterogeneous setup, they reached up to 90\% accuracy on both datasets.
While we confirmed that our attack is well implemented, a key question appears: \textbf{what causes these contradictory results?}

\paragraph{Alternative non-member sampling}
The main difference between our results and \cite{humphries_investigating_2023} resides in the non-member sampling.
While we sample the non-member from the same distribution as the target's dataset, Humphries et al. \cite{humphries_investigating_2023} sampled them from a third distribution (different from the attacker's and target's).

\begin{figure}[t]
    \centering
    \includegraphics[width=.8\linewidth]{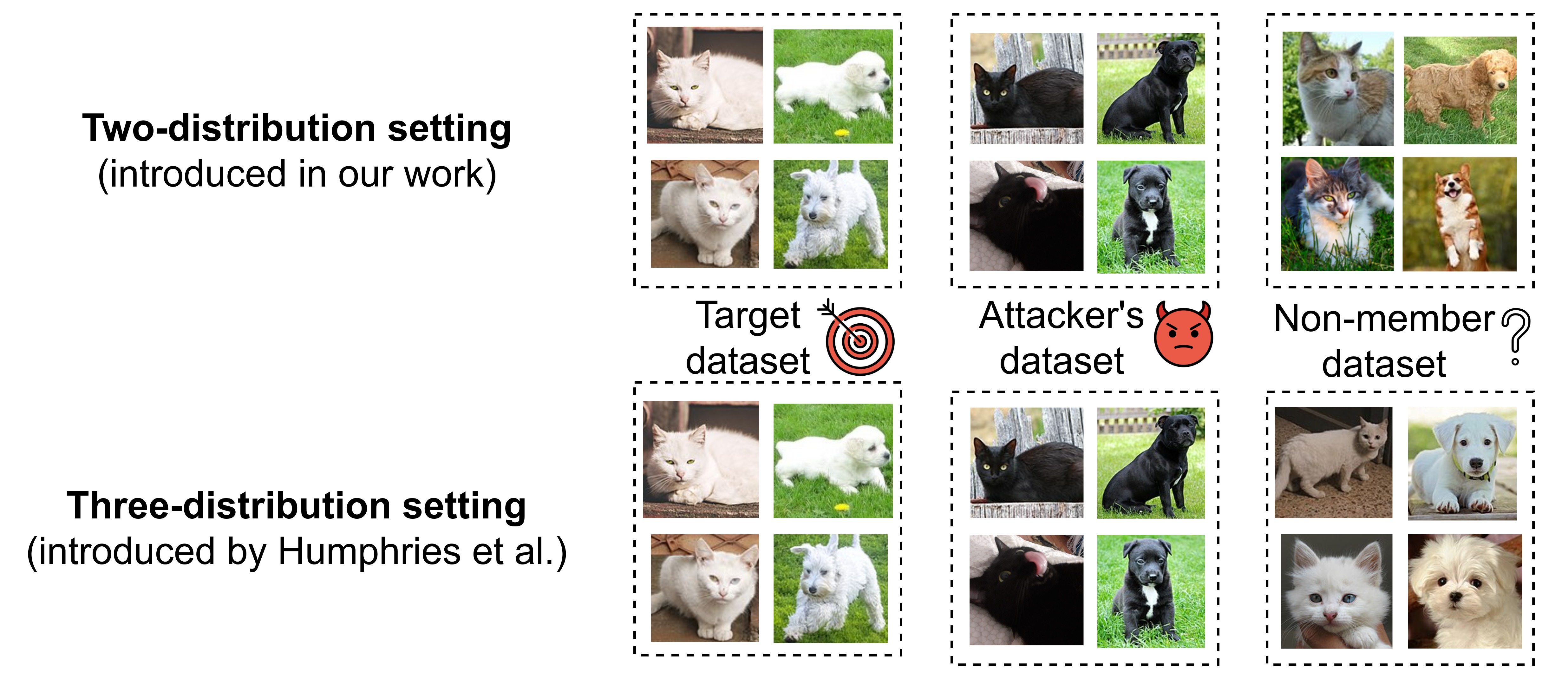}
    \caption{Approaches to sample heterogeneous data in MIA}
    \label{fig:non-member-sampling}
\end{figure}

Figure \ref{fig:non-member-sampling} illustrates this difference using an animal image dataset.
In this simplistic example, using our sampling, both the target and non-members data would be white animals and the attacker's dataset would be black animals.
Using \cite{humphries_investigating_2023}, the target would be white animals, the attacker would have black animals, and the non-members would be animals of multiple colors.

While the difference seems subtle, these sampling methods produce two distinct attack challenges.
In our case, the attacker must identify which white animals were part of the dataset.
It requires \emph{identifying the individuals}.
In \cite{humphries_investigating_2023}, the attacker simply infers whether an individual belongs to the same distribution as the training data.
It requires \emph{identifying the distribution}.
The attacker in \cite{humphries_investigating_2023} does not really identify specific white cats (like in our attack), but simply needs to infer that the target model was trained on white cats.
In this sense, \textbf{their attack setup could be interpreted as a ``distribution membership'' inference attack}.

\begin{table}
    \centering
    \begin{tabular}{|c|c|c|}
        \hline
        {Sampling}                                                           & {Average Accuracy} \\
        \hline
        \hline
        2-distribution setting                                               & 47.75              \\\hline
        25\% of non-members from 3rd distribution                            & 60.71              \\\hline
        50\% of non-members from 3rd distribution                            & 57.78              \\\hline
        75\% of non-members from 3rd distribution                            & 70.24              \\\hline
        3-distribution seeting (like in \cite{humphries_investigating_2023}) & 91.23              \\\hline
    \end{tabular}
    \caption{MIA accuracy on the ``\texttt{Heart}'' dataset for varying non-member sampling.}
    \label{tab:mia-non-members}
\end{table}

\paragraph{From zero to hero}
Table \ref{tab:mia-non-members} presents the MIA accuracy for a varying proportion of non-members drawn from a third distribution: 0\% corresponds to our 2-distribution setting, and 100\% corresponds to the 3-distribution setting of \cite{humphries_investigating_2023}.
This table confirms that the more non-members are sampled from a third-party distribution, the higher the MIA accuracy is.
With 3 distributions, we obtain results similar to those reported in \cite{humphries_investigating_2023}.
Thus, \textbf{the non-member sampling was the cause of the contradictory results.}

\section{Conclusion}
Our work introduced novel tools to evaluate MIA in heterogeneous-data environments.
On the one hand, we proposed a heterogeneity metric usable on any tabular dataset.
On the other hand, we compared two sampling methods to simulate MIA in heterogeneous-data setups.
Our experiments showed that the subtle differences in these setups lead to seemingly contradicting results: high attack accuracy in one setting and insusceptible for MIA in the other.

\paragraph{Future works}
We lack one uniform theoretical model for MIA in heterogeneous-data setups.
While classic MIA setups are modeled using a single data distribution \cite{chatzikokolakis_bayes_2023}, data heterogeneity raises a theoretical problem: should our theoretical model include two distributions (i.e., target and attacker), three (i.e., target, attacker, and non-member), or even more (e.g., if the attacker owns multiple datasets from different distributions)?
Further theoretical work is necessary to formalize and standardize MIA in heterogeneous-data setups.
Such theoretical work should be considered a \emph{generalization} of existing MIA works.

Far from being only a theoretical discussion, the non-member sampling has major practical impacts.
For example, with three-distributions, the attacks are much stronger, so attack mitigation (as developed in \cite{humphries_investigating_2023}) is mandatory.

Finally, our work provided first experimental results that need to be extended to other attacks and non-tabular (but more complex) data types.

\begin{credits}
    \subsubsection{\ackname}
    This work is based on the MSc thesis of Bram van Dartel \cite{dartel2024effect}, and was supported by the Netherlands Organization for Scientific Research (De Nederlandse Organisatie voor Wetenschappelijk Onderzoek) under NWO:SHARE project [CS.011].
\end{credits}
%
%

\bibliographystyle{splncs04}
\bibliography{ref}

\end{document}